\documentclass{elsart}

\usepackage{epsfig}
\usepackage{amssymb,amsmath}

\begin{document}
 
\begin{frontmatter}          
 
\title{Localized defects in a cellular automaton model for traffic
flow with phase separation}

\author[Duisburg]{A. Pottmeier}, 
\author[Duisburg]{R. Barlovic},
\author[Duisburg]{W. Knospe},
\author[Cologne]{A. Schadschneider}, and
\author[Duisburg]{M. Schreckenberg}
 
\address[Duisburg]{Theoretische Physik (Fakult\"at 4),
        Gerhard-Mercator-Universit\"at Duisburg,
        D-47048 Duisburg, Germany,
        email: pottmeier,barlovic,knospe,schreckenberg@uni-duisburg.de}
\address[Cologne]{Institut f\"ur Theoretische Physik,
        Universit\"at zu K\"oln,
        D-50937 K\"oln, Germany,
        email: as@thp.uni-koeln.de}
                                        
\begin{abstract}
We study the impact of a localized defect in a cellular
automaton model for traffic flow which exhibits metastable states and
phase separation. The defect is implemented by locally limiting
the maximal possible flow through an increase of
the deceleration probability.
Depending on the magnitude of the defect three phases can
be identified in the system. 
One of these phases shows the characteristics of stop-and-go traffic 
which can not be
found in the model without lattice defect. Thus our results provide
evidence that even in a model with strong phase separation stop-and-go 
traffic can occur if local defects exist. 
From a physical point of view the model
describes the competition between two mechanisms of phase separation.
\end{abstract}
\end{frontmatter}


\section{Introduction}
\label{introduction}
Nowadays mobility, mostly realized by means of vehicular traffic, is an
integral part of any modern society. Unfortunately, the
capacity of the existing road networks is often exceeded in densely
populated areas and the possibilities for a further expansion 
are very limited. Therefore experimental and theoretical
investigations of traffic flow have been the focus of extensive
research during the past decades with the aim to optimize the usage of
existing capacities. 

In many cases it seems as if the occurrence of traffic
states with high density like jammed or synchronized traffic
can be linked to external influences, e.g., on- and off-ramps,
bottlenecks, lane reductions or road works
(see~\cite{daganzo,kerner96,kerner97,kerner98,helbing1,neubert,lee,popkov}).
Such local perturbations can act as a seed for the formation of
jammed traffic states which finally spread through large parts of
the network. Hence local defects can have a crucial impact on the
overall network performance. A proper understanding of the interactions
between local defects and model dynamics and its consequences 
is indispensable for realistic large scale computer simulations.

In recent years various theoretical approaches for the description
of traffic flow have been suggested (for reviews, see
e.g,~\cite{tgf95,tgf97,tgf99,review,helbing2}) which
can be divided into macroscopic and microscopic models. 
Today one of the main interests in applications to
real traffic is to perform real-time simulations of large networks
with access to individual vehicles. 
In contrast to macroscopic models, microscopic models provide
information about the behavior of individual vehicles.
Due to their simplicity cellular automata (CA) have become quite
popular in the field of microscopic traffic modeling. The first CA
model for traffic flow with the ability to reproduce the basic
phenomena encountered in real traffic, e.g., the occurrence of phantom
traffic jams, was proposed by Nagel and Schreckenberg
(NaSch)~\cite{nasch}. However, the NaSch model can by far not explain
all empirical results like the velocity of jam fronts or the
reduced outflow from jams leading to metastable states. Therefore
modifications have been suggested to obtain a description of the
traffic dynamics on a more detailed level. In this paper we focus our
investigations on the NaSch model with velocity-dependent
randomization (VDR model)~\cite{barlovic1} which exhibits metastable
states and shows phase separation into free flow and wide
jams. Because of the strong phase separation it turns out that the VDR
model is an interesting candidate for investigating the influence of
external perturbations on the internal model dynamics. Due to the fact
that only wide jams appear in an undisturbed system, any additional
high density patterns induced by external forces can be identified
easily.

The impact of defects in the NaSch model (and related models, e.g.,
the asymmetric simple exclusion process (ASEP)) is by now well
understood. Basically two types of defects can be distinguished which
can be characterized as particlewise and sitewise disorder, respectively
\cite{krug99}. In a model where a finite number (in the thermodynamic
limit) of particles (vehicles) or sites has different properties from
the rest these are usually called {\em defects}. In the first case,
corresponding to a limiting case of particlewise disorder
\cite{evans,Krug96,Krug97}, the defect particles may have a smaller 
maximal velocity. Such defects are not localized in space in contrast to 
those corresponding to sitewise disorder where in a localized region
certain parameters of the model take different values, e.g., by 
imposing a speed limit or increasing the deceleration probabilities 
\cite{lebo,YKT,vicsek,emmer,defect,NagataniD,santenpromo,pottidiplom}.

In both cases a parameter regime exists where the global behavior
of the system is controlled by the defect which acts as a bottleneck. 
Generically it induces phase separation into a high and a low density
region separated by a sharp discontinuity (``shock'').
In the case of particlewise disorder with one slow car (and no overtaking)
the faster cars tend to pile up behind the slow one. This behavior
has certain similarities with Bose-Einstein condensation \cite{evans}.
For a spatially localized defect one also finds a separation into a
high and a low density regime, but with the high density region pinned
to the defect. This behavior has been found in a variety of models and
for different defect realizations.
However, none of the models investigated so far exhibits phase
separation through the existence of high-flow metastable states
which is an important ingredient for any realistic traffic model.

In the present paper we investigate the influence of a localized defect
on a traffic model exhibiting metastable states. This allows to
study the interplay between two very different mechanisms for phase
separation, i.e., one driven by the dynamics of the particles and
one driven by the defect. Our investigations of the VDR model reveal 
the occurrence of three different phases
whereby one of them shows the characteristics of stop-and-go
traffic. Note, that these phases can not be obtained in the model
without lattice defect. Moreover, concerning the question whether
certain jam patterns found in real traffic are induced by local
defects or due to the internal behavior of the drivers our findings
allow a deeper insight into possible methods of modeling such traffic
states. In the VDR model individual driving behavior is realized by a
simple stochastic parameter. For the purpose of keeping the set of
model parameters manageable we implemented the local defect in our
investigations simply by increasing this stochastic noise in a
specific part of the system. 

\section{Definition of the Model}
\label{model}

\begin{center}
\begin{figure}[h]
 \centerline{\epsfig{figure=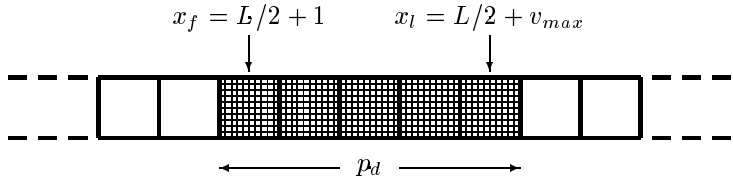}}
 \caption{\protect{Schematic representation of the local defect. The defect
itself is placed at a fixed position on a periodic one lane street. Its width 
is chosen to $v_{\text{max}}$ cells.}}
\label{defect}
\end{figure}
\end{center}

In the spirit of modeling complex phenomena in statistical physics one
is interested in keeping the model as simple as possible. Obviously,
the modeling of traffic flow in the language of cellular automata (CA)
always implies an extreme simplification of real world
conditions. Hence space, speed, acceleration, and even time are
treated as discrete variables and the motion is realized in the ideal
case by a minimal set of local rules. In this manner we tried to find
a straightforward representation for local defects on a one-lane
street in the VDR model. Since the model contains a stochastic
parameter which is needed to implement various phenomena found in real
traffic, e.g., spontaneous jam formation, reduced outflow from a jam,
it seems obvious to implement a local defect by increasing this
parameter in a limited area. As mentioned before, other types of
localized defects have been studied in the NaSch model and related
models, e.g., reducing the maximal velocity of the cars
locally. Introducing a locally lower speed limit in the VDR model
should have the same effect as the enhanced deceleration probability
since one expects also phase separation into high and low density
regions in a certain regime of the global density. However, the
microscopic details of the states can be different for the various
defect types. 

For the sake of completeness, we will briefly recall the definition of
the VDR model~\cite{barlovic1}. The road is divided
into cells of length $7.5m$. Each cell can either be empty or occupied
by just one car. The speed of each vehicle can take one of the
$v_{\text{max}}+1$ allowed integer values $v=0,1,...,v_{\text{max}}$. 
The state of the road at time $t+1$ can be obtained from that at
time-step $t$ by applying the following rules to all cars at the same time 
(parallel dynamics):
 
\begin{itemize}
\item Step 1: Acceleration:\\
$v_{n}\rightarrow \min(v_{n}+1,v_{\text{max}})$\\[0.2cm]
\item Step 2: Braking:\\
$v_{n}\rightarrow \min(v_{n},d_{n}-1)$\\[0.2cm]
\item Step 3: Randomization: with probability $p = p(v)$\\   
$v_{n}\rightarrow \max(v_{n}-1,0)$\\[0.2cm]
\item Step 4: Driving:\\
$x_{n}\rightarrow x_{n}+v_{n}$
\end{itemize}
 
Here $d_{n}=x_{n+1}-x_n$ denotes the distance (`headway') to the next car 
ahead. 
One time-step corresponds to approximately $1s$ in real time. Note, that 
the fluctuation parameter $p = p(v)$ depends on the velocity whereas it 
is constant in the NaSch model.
This parameter has to be determined before the acceleration in ``step 1''.
For simplicity we study the so called slow-to-start case with
two stochastic parameters:
\begin{equation}
p(v) = \left \{
        \begin{array}{ccc}
        p_{0} & \qquad\text{for} & v = 0, \\
        p & \qquad\text{for} & v > 0.
   \end{array}
        \right.
\label{pvonv}
\end{equation}
Thereby, $p_0$ controls the fluctuations of cars a rest, i.e.,
determines the velocity of a jam, and $p$ controls the velocity
fluctuations of moving cars. In the case $p_0 \gg p$ the model shows
the expected features: phase separation and metastable states.
It is interesting that an alternative choice of $p(v)$, e.g., $p_{0} \ll
p$ leads to a completely different behavior. Note that
for $p_{0} = p$ the original NaSch model is recovered. If not stated
otherwise, $p_{0}$ is set to $0.5$ and $p$ to $0.01$ throughout this paper.
These parameter values yield a good agreement with empirical results 
(see~\cite{barlovic1,barlovic2} for further details).

As mentioned before the local defect is implemented by increasing the
stochastic noise parameter $p(v)$ in a limited area of the system.
Various implementations of a local defect are possible,
but here we want to keep the set of parameters as
small as possible. The length of the defect is chosen to
$L_d$ cells and the stochastic noise $p(v)$ is replaced by a defect
noise $p_d$. The defect length $L_d$ itself is set to $L_d = v_{\text{max}}$
to ensure that each car will participate at least once in an update
with the enhanced breaking probability $p_d$. 
Note, that this also implies that slow vehicles 
underly a stronger influence. They need more than one timestep to
cross the defect region and thus the defect deceleration rule has
to be applied more often than for fast cars which can cross the
defect in one timestep.

Given that the stochastic noise in the VDR model depends on the velocity 
of vehicles, the choice of the stochastic parameter inside the defect
must be seen with respect to this. Our strategy is to choose 
the stochastic noise in a way that it is maximal. In detail one gets
the following equation:           
\begin{equation}
p(x,v) = \left \{
        \begin{array}{ccc}
        \max(p(v),p_{d}) \qquad  & \text{for} & x \in D, \\
        p(v) \qquad & \text{for} & x \notin D,
   \end{array}
        \right.
\label{pd}
\end{equation}
with $D = \{x | x_f \le x \le x_l\}$ denoting the cells forming 
the defect. Here $x_f$ is the first cell of the defect and $x_l$ the
last one. In Fig.~\ref{defect} a schematic representation of the local
defect with all its parameters is depicted. Obviously, the defect
width as well as $p_d$ control the strength of the defect. For the
sake of simplicity, we keep the width of the defect fixed and chose
$p_d$ as control parameter.
The influence of different spatial extensions of the defect
is studied in~\cite{pottidiplom}.


\section{Numerical results}
\vspace{0.5cm}
\label{results}
\begin{figure}[h]
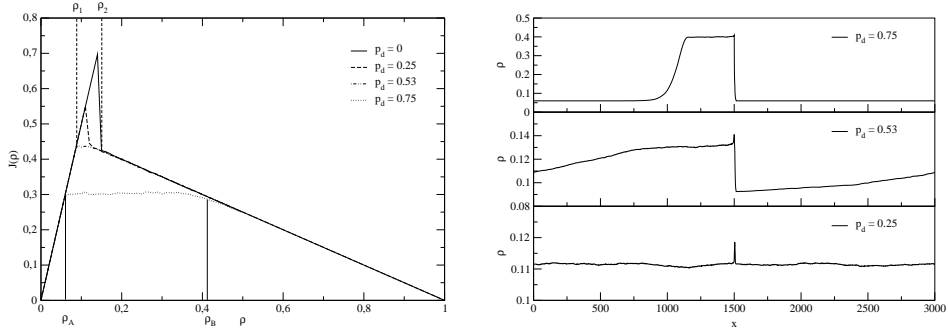

 \centerline{\psfig{figure=VDR_Defect_Fund-1.eps,width=6cm}
\quad\psfig{figure=density_paper-1.eps,width=6cm}}
\caption{\protect{ {\bf Left:} Typical fundamental diagram of the VDR
model with a lattice defect. The different defect noise parameters
$p_d$ cover the occurring phases. The remaining model parameters are
$L=3000$, $v_{\text{max}}=5$, $p_0=0.5$ and $p=0.01$. 
{\bf Right:} Density profiles of the analyzed system for
density $\rho=\frac{1}{9}$. Starting from bottom to top: The case
$p_d=0.25$ corresponds to the VDR phase.
Only a small peak
at the defect is observable representing the interactions between the
defect and the vehicles. For an intermediate defect noise, here $p_d=0.53$, 
the density profile is almost linear due to the different lifetimes of the
small jams emerging at the defect. In the high defect noise case with
$p_d=0.75$ (stop-and-go phase) the system self-organizes into macroscopic
high and low density regimes similar to the NaSch model with defect.}}
\label{fundi}
\end{figure}            
In this section we discuss the impact of the local defect scenario
introduced in the previous section on the basis of numerical results.
We use the stochastic noise $p_{d}$ inside the local area belonging to
the defect as control parameter. The other parameters of the model are 
kept fixed ($L=3000,v_{\text{max}}=5,p_0=0.5,p=0.01$). Starting from 
low $p_d$ three different phases can be distinguished in the system.

At this point it should be mentioned that the fundamental diagram of
the VDR model shows three different regimes with respect to the
density (Fig.~\ref{fundi}, $p_d=0$). Up to a certain density
$\rho_{1}$ the system is in a free flow state, i.e., no jams exist.
Above the density $\rho_{2}$ the system resides in the so called
jammed state. This state is characterized by wide phase separated
jams and free flowing vehicles. However, the most interesting regime
lies between the two densities $\rho_{1}$ and $\rho_{2}$ where the
system can be in two different states. One is a metastable homogeneous
state with an extremely long lifetime and a high flow where jams can
appear due to internal fluctuations. The other one is a phase
separated state with wide jams which can be reached through the decay
of the homogeneous state or directly owing to the initial
conditions. A typical fundamental diagram for the considered system is
plotted in Fig.~\ref{fundi} for different $p_d$.  The case $p_d=0$
corresponds to the undisturbed VDR model. The metastable branch with
high flow between $\rho_{1}$ and $\rho_{2}$ can clearly be identified.


\subsection{Small defect noise: $p_d \ll p$}

For a low stochastic noise inside the defect $p_d \ll 1$, the
influence on the overall dynamics of the system is negligible.
The only exception is the lifetime of the metastable states occurring 
in the VDR model which is decreasing strongly. The reason
is that the metastable states of the VDR model are very sensitive
in regard to perturbations. In~\cite{barlovic3} an analytical expression for
the sensitivity, i.e., the probability that a perturbation of finite
magnitude destroys the metastable state, is presented. As one can see
in Fig.~\ref{fundi} (left), the maximal possible flow of the
metastable branch is reduced extremely for $p_{d}=0.25$ which is
chosen as a typical value for a small defect noise. Obviously other
values for $p_{d}$ lead to different lifetimes. However, the term
``small noise'' should not be related to the lifetimes of the
metastable states but rather to the impact on the systems dynamics
after the transition from a metastable high-flow state to a jammed
state. Thus a defect noise is ``small'' if the jammed state of the
system is nearly unaffected by it. This phase will be denoted as the
{\em VDR phase} since it matches with the jammed state of the VDR
model. It is characterized by a wide jam, which moves backwards and is
able to pass the defect area uninfluenced, and free flowing
vehicles. The distribution of headways in the latter is determined
completely by the outflow from the jam.
The distance between the free
flowing vehicles is large enough to absorb additional velocity
fluctuations in the area of the defect without the emergence of new
jams. In Fig.~\ref{fundi} (right) the density profile of the
considered system is shown for various $p_d$. The VDR phase
shows a constant density profile with a small peak in the area
of the defect. This peak is created by the additional velocity
fluctuations which lead to increased travel times. Besides this peak
there is no markable difference to the jammed state of the VDR model.


\subsection{Large defect noise: $p_d \gg p$}

\begin{figure}[h]
 \centerline{\psfig{figure=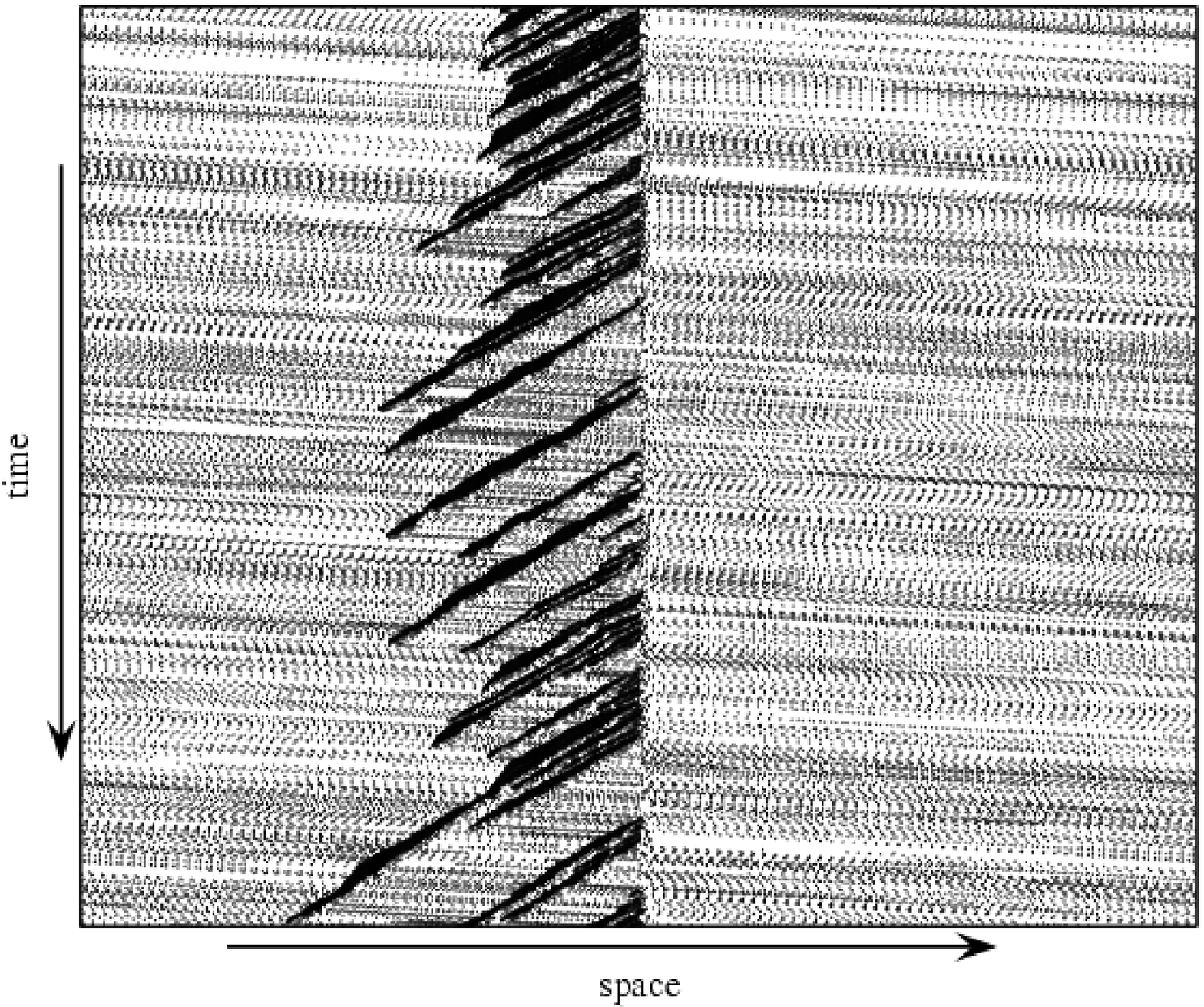,width=6cm}
\quad\quad
\psfig{figure=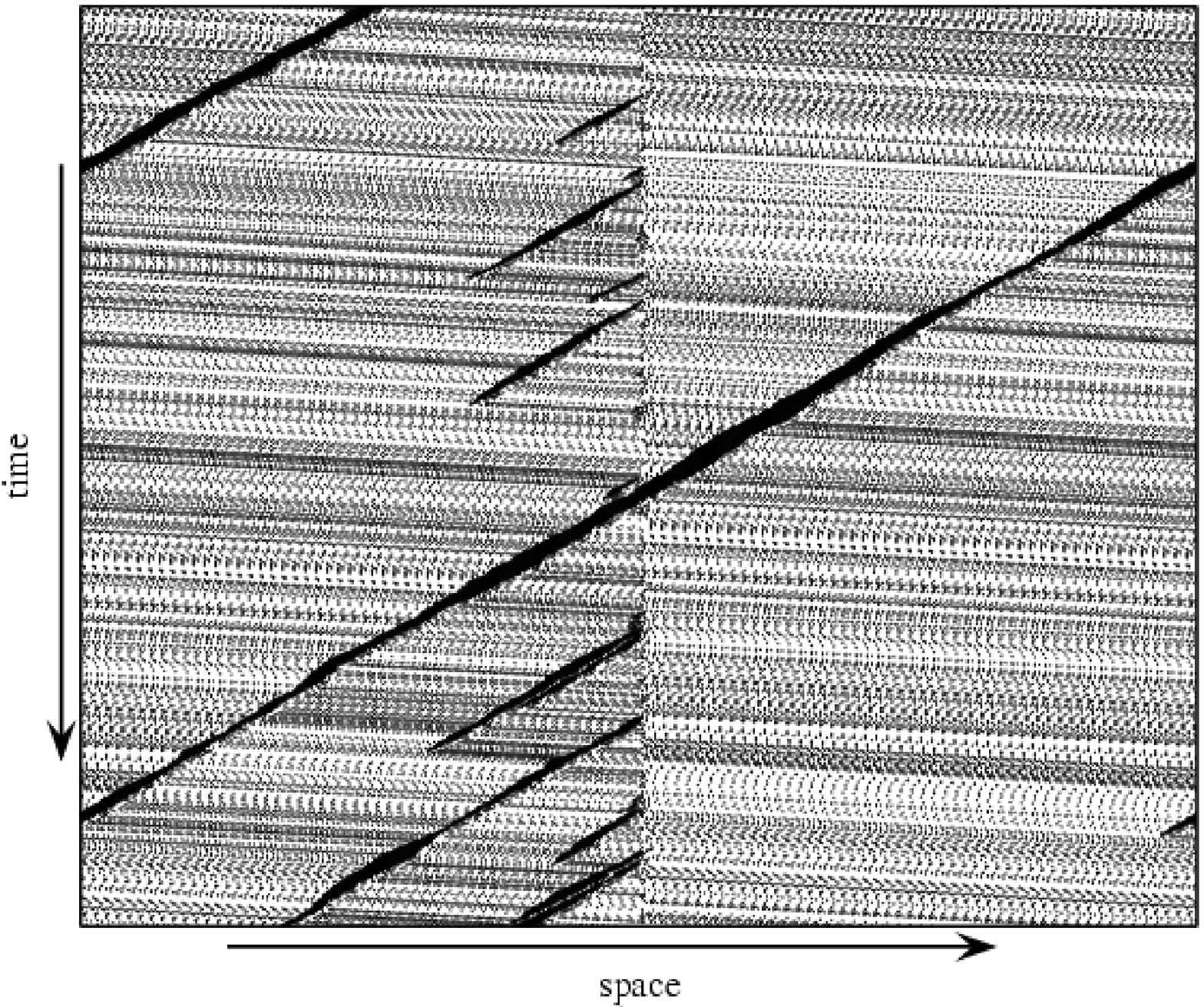,width=6cm}}
\caption{\protect{Space-time plot of the analyzed system for a density
of $\rho=\frac{1}{8}$.  {\bf Left:} The high defect noise ($p_d=0.75$)
state is characterized by a high density region consisting of small
compact jams which are separated by small free flow regions. Whereas
this jam pattern is pinned at the defect and shows some similarities
to stop-and-go traffic. {\bf Right:} For an intermediate defect noise
$p_d=0.53$ a mixture of a wide jam moving nearly undisturbed through
the system and small jams which are formed at the defect is
observable. The different lifetimes of the small jams lead to an
approximately linear
density profile (see Fig.~\ref{fundi} right).}}
\label{spacetime}
\end{figure}

As expected, large defect noises $p_d$ have a significant influence
on the flow of the system. Three different density regimes can be 
distinguished corresponding to the ones found in the NaSch model under the 
same circumstances~\cite{santenpromo}. At low densities the average distance
between the vehicles is large enough to compensate velocity
fluctuations induced by the defect. Similarly for high densities the
system is dominated by jams whose movement is nearly unhindered. Thus
the fundamental diagram in Fig.~\ref{fundi} coincides for these two
density regimes with the one of the undisturbed model.  However, the
most interesting density regime is situated in the middle of the fundamental 
diagram and can be identified by a plateau. This plateau is formed since 
the capacity of the defect limits the global flow in the system. It can 
not exceed the maximal flow $J_{\text{def}}$ through the defect which 
therefore cuts off the fundamental diagram at $J_{\text{def}}$ and leads 
to the formation of the plateau.
The plateau value decreases almost linearly with an increasing defect
noise.  As one can see in the corresponding space-time plot
Fig.~\ref{spacetime} (left), a considerable amount of vehicles is
gathered at the defect forming a high density region. The width of
this high density region itself grows linearly with increasing density
as long as densities corresponding to the plateau are considered
$\rho_A < \rho < \rho_E$ (for a detailed analysis
see~\cite{pottidiplom}). Fig.~\ref{fundi} (right) shows the density
profile for a density within the plateau region.  The system
self-organizes into a macroscopic high density region pinned at the
defect and a low density region determined by the capacity of the
defect.  So far the macroscopic properties are comparable to results
obtained by the NaSch model with a local defect~\cite{santenpromo}.
However, a look at the microscopic structure of the high density
region reveals some interesting differences. In contrast to the NaSch
model, where the high density region at the defect consists of a
compact congested region, the high density region in the VDR model is
characterized by small compact jams which are separated by free flow
regimes (see Fig.~\ref{spacetime} (left)). The term ``small compact
jams'' means that the jams at the defect are significantly smaller
than the width of the high density region.
This specific jam pattern shows some similarities with stop-and-go
traffic and will therefore be called {\em stop-and-go}
phase\footnote{Note, that several congested traffic states, e.g.,
synchronized traffic, can exist in reality (see~\cite{tgf99} for an
overview).}. However, the correspondence to real world traffic
patterns should be viewed in a qualitatively sense here. The {\em stop
and go} phase is characterized by a relatively large jammed region
(high density) consisting of jams alternating with free flow
sections. Here it is important to keep in mind that the high density
region is not distinguishable from the one occurring in the NaSch
model if considering only macroscopic quantities. The microscopic
structure shows a new high density state which cannot be found in the
NaSch model. Therefore it will be analyzed in a future work in more
detail. Moreover, it must be stressed that the local defect is an
essential ingredient for the occurrence of stop-and-go traffic in the
VDR model (an undisturbed system shows only free flow or one single
wide jam) while in the NaSch model jams of various sizes can occur
even in an undisturbed system~\cite{review}. One has to take into
account that the so called slow-to-start case $p_0 \gg p$ of the VDR
model is considered here, which leads to the occurrence of compact 
wide jams. 
In general for parameter combinations with $p<p_0$ even in the VDR model 
stop-and-go traffic can be found.
Thus the absence of this specific traffic state is not a limitation
of the model, but rather due to the special choice of fluctuation 
parameters. 
However, we have focussed on the slow-to-start case $p_0 \gg p$ here
since we are interested in the effects of the competition between the
bulk and the defect dynamics.
\\[0.2cm]

\subsection{Transition regime: $p_d \approx p$}

In the following we focus our investigations on intermediate values of
the defect noise parameter $p_d$. The space-time plot in
Fig.~\ref{spacetime} (right) exhibits the microscopic structure of the
stationary state for this parameter region. In particular a wide jam
moves backwards through the system and additionally some small jams
are formed at the defect which have a limited lifetime. Note, that as
in the case of large and small $p_d$ the defect has almost no
influence on the dynamics of the system for low and high
densities. Therefore we will focus on intermediate densities to
demonstrate effects caused by the additional localized noise parameter
$p_d$. As one can see in Fig.~\ref{fundi} the described mixture of a
wide jam and some small ones cannot be identified easily in the
fundamental diagram. The flow is almost identical to that in the model
without defect except for the missing metastable branch.  This is
rather different from the case of large $p_d$ where a plateau is
formed in the fundamental diagram. This suggests that for an
intermediate $p_d$ the capacity of the defect is close to the maximum
possible flow in the system. Furthermore, taking a look at the density
profile for a corresponding intermediate defect noise (respectively
$p_d=0.53$) another difference to the large $p_d$ case is
observable. For a large $p_d$ the system self-organizes into a
macroscopic high density region pinned at the defect and a low density
region. In contrast, for intermediate values of $p_d$ the density
profile decreases approximately linear in upstream direction at the
defect. This behavior can be traced back to the different lifetimes
of the small compact jams. The dynamics of such a small jam can be
described analytically by random walk arguments as shown
in~\cite{barlovic3}.

This marks an important difference to the behavior of the NaSch
model with a defect. Here a comparable state does not exist.
This finding is also interesting for the interpretation of empirical results.
Traffic states consisting of wide jams passing a localized region with a 
flow comparable to free flow and small mean velocity (these states are 
often denoted as synchronized traffic) are observable in real
traffic~\cite{kernerneu1,kernerneu2}.
However, the system state for intermediate $p_d$ has to be
interpreted as a crossover phase. For small $p_d$ it was shown
that only one single wide jam moves undisturbed through the system
while for a high $p_d$ no single wide jam can exist. In contrast a region
with many small jams is formed at the defect (stop-and-go
traffic). Starting from small $p_d$ without any small jams in the
system one can observe the occurrence of small jams at the defect with
an increasing frequency if increasing the defect noise $p_d$. 
Further increasing the defect noise finally leads to the complete
dissolving of the large jam. Now the system shows only
stop-and-go traffic in the vicinity of the defect.
To identify this transition between the crossover phase
containing one large and various small jams and the stop-and-go phase
(large $p_d$) we investigate the autocorrelation function in the
following.\\
\begin{figure}[h]
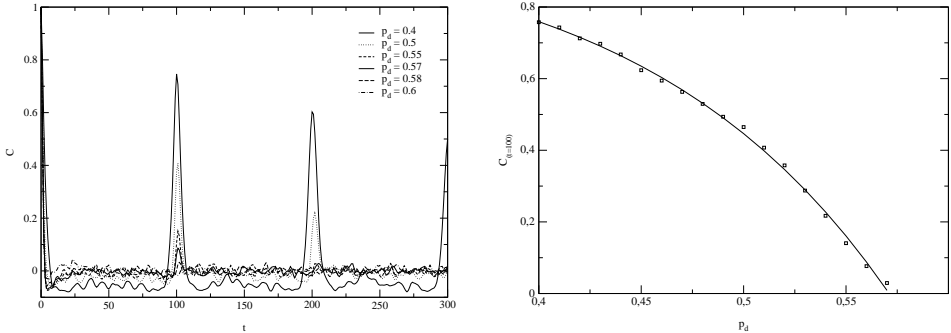

 \centerline{\psfig{figure=autokorrelation.eps,width=6cm}
\quad
\psfig{figure=ausgleich_fit.eps,width=6cm}} \caption{\protect{ {\bf
Left:} The autocorrelation function shows peaks at a regular distance
representing the wide backward moving jam in the system. The height of
the peaks decreases with increasing $p_d$ until the peak vanish
completely because the transition to the stop and go phase occur.
{\bf Right:} Plot of the maximum of the first peak as function of the
defect noise $p_d$. The vanishing point of the peak is determined 
through an extrapolation is plotted using an exponential fit function.}}
\label{auto}
\end{figure}         

The density autocorrelation function $\mathcal{C}$ which is defined
as\footnote{The measuring time interval $T$ is set to $T = 60$
timesteps in our simulations.}
\begin{equation}
\mathcal{C} = \frac{1}{T} \frac{\sum_{\tau = 1}^T \rho_{l_\tau}
\rho_{l_{\tau + \Delta T}} - \frac{1}{T} \sum_{\tau = 1}^T
\rho_{l_\tau} \sum_{\tau = 1}^T \rho_{l_{\tau + \Delta T}}}{\langle\rho^2
\rangle-\langle\rho\rangle^2},
\label{def_autocorr}
\end{equation}
with the local density $\rho_l$ of a site at position $x_n$,
\begin{equation}
\rho_l (i) = \frac{1}{T} \sum_{t=1}^T \frac{1}{v_n}\;\; \qquad\quad
\mbox{with}\; (x_n - v_n) < i\;\; \mbox{and}\;\; x_n > i,
\end{equation}
can be used to 
distinguish between the phase with stop-and-go traffic (pinned at
the defect) and the intermediate $p_d$ region where crossover behavior
is observed. The autocorrelation function is able to
detect periodically moving structures, i.e., large moving jams. 
For small $p_d$ (VDR phase) as well as for intermediate
$p_d$ (crossover phase) the system is characterized by a wide jam
which recurs periodically due to the periodic boundary conditions.
The autocorrelation function is capable to specify the point at which
this wide jam vanishes, i.e., the transition point to the pinned 
stop-and-go phase. Remember, that in the case of crossover behavior
additionally small jams are formed at the defect. However, these small
jams do not affect the autocorrelation function at all since their
lifetime is not large enough to move throughout the whole system. 

%
%
%

In Fig.~\ref{auto} (left) the density autocorrelation function is
plotted against the time lag $\Delta T$. The curve shows peaks with a
regular distance which represent the wide jam moving backwards
through the periodic system. It can clearly be seen that the height
of the peaks decreases with increasing $p_d$ until they finally vanish
completely, i.e., the wide jam dissolves and the transition to the
pinned phase occurs. Besides, the autocorrelation function can be used
to determine the velocity of jams in the model which is given by the
distance between the peaks and the corresponding length of the system
(see~\cite{neubert2} for further details).

In order to determine the transition point we concentrate on the 
absolute value of the first peak. This peak value is plotted as
function of the defect noise parameter $p_d$ (Fig.~\ref{auto}). 
To obtain the transition
point the curve is extrapolated to zero, i.e., vanishing peak values.
This is done using the fit function $f(x) = 1-\alpha e^{-\beta x}$ which
reproduces the shape of the curve quite accurately. In
Fig.~\ref{auto} (right) the values of the first peak are shown 
with the corresponding exponential fit for the global density $\rho =
1/6$. For these parameters we get the best fit with $\alpha = 0.01$ and
$\beta = -7.9$. The crossing through zero is obtained for a defect
noise of $p_d = 0.57$. For further results concerning the transition
to the pinned phase see~\cite{pottidiplom}.

\subsection{Relevance for systems with ramps}

It has been realized recently \cite{Lee98,Lee99,Helbing99,popkov} 
that inhomogeneities like on- and off-ramps play an important role in 
real traffic. They might be the origin of a variety of different traffic 
states observed empirically.
For the NaSch model it has been found in \cite{diedrich} that the
effects of ramps are very similar to those of localized defects.
The presence of an on-ramp leads to a local increase of the density
and a restriction of the maximal possible flow. 
The fundamental diagram shows a plateau and the plateau value is
determined by the inflow from the ramp. Also the microscopic structure
of the states in systems with defects and ramps is very similar.

For the case of the VDR model with ramps we also find for small
ramp flows a phase similar to the VDR phase in defect systems.
However, the width of the jam varies close to the on- and off-ramp
where it becomes larger or smaller, respectively.
For large ramp flows a phase similar to the stop-and-go phase
in the defect model is realized. It is characterized by a high density
region of stop-and-go traffic pinned to the ramp. 
Furthermore a transition region can be identified where a large moving
jam coexists with smaller jams which are pinned at the ramps.

This shows that also the VDR models with defects and ramps exhibit
a rather similar behavior. However, there are subtle differences
which are discussed in \cite{pottidiplom}.


\section{Summary and discussion}
\label{conclusion}

We have analyzed the impact of a local lattice defect in the VDR model.
An important aspect is the competition between two mechanisms of
phase separation. The dynamics of the VDR model leads to a phase 
separation at high densities into a large moving jam and a free flow
region. In contrast, a localized defect triggers the formation of
a high density region pinned at the defect.
From a practical point of view our aim was to
obtain a deeper insight into the formation of jam patterns due
to topological peculiarities. The local defect itself was implemented
by increasing the stochastic noise of vehicles within a
certain area.
Three different system states (phases) can be observed as the
defect noise $p_d$ is varied. Small defect noises $p_d$ 
reduce the lifetimes of the metastable states in the VDR model which
show a strong sensitivity to disturbances. The vehicles in the jammed 
state of the system, consisting of a single wide jam and free flow, can 
pass nearly undisturbed through the defect. We denoted this
phase as {\em VDR phase} since there is almost no difference to the jammed
state of the VDR model without defect. In contrast to the low $p_d$
case, for a large $p_d$ a pinned high density region is formed at the
defect limiting the overall system flow. The microscopic structure of
this high density region reveals the occurrence of small compact jams
which are separated by small free flow regions. This phase is
called {\em stop-and-go phase} since the jam pattern shows strong 
similarities to stop-and-go traffic. An important point is that
stop-and-go traffic cannot be found in the VDR model without a lattice 
defect. Furthermore, we found crossover behavior for an intermediate
defect noise $p_d$. Here a wide jam moves backwards through
the system. Additionally small jams are formed at the defect which have 
a limited lifetime. In order to determine the transition point between
this crossover phase, containing one wide jam and various small jams,
and the stop-and-go phase, where no wide jam can be found,
we investigated the density autocorrelation function. 

To conclude, the results presented here are of practical relevance for
traffic flow simulations using simple stochastic CA models like the
VDR model. Complex networks usually contain many defects such as
crossings, lane reductions, on- and off-ramps, a detailed
understanding of their influence is of importance.  Furthermore
it is very interesting that system states like stop-and-go traffic
can emerge through the introduction of defects even if they can not
be realized by the dynamics of the model alone.
Concerning the question whether certain system states, like 
stop-and-go traffic or synchronized traffic, found in reality are 
induced by topological aspects or the drivers behavior our findings 
could be very benefitable since they imply a strong influence of 
defects.

\end{document}